\documentclass[journal,english,10pt,letterpaper]{IEEEtran}
\usepackage{babel}
\usepackage[utf8]{inputenc}
\usepackage{graphicx}
\usepackage{subfigure}
\usepackage{amsmath}
\usepackage{url}
\usepackage{color}\usepackage{tikz}
\usepackage[colorlinks,bookmarksopen,bookmarksnumbered,citecolor=red,urlcolor=re
d]{hyperref}

\newcommand\copyrighttext{%
  \footnotesize \textcopyright 2008 IEEE. Personal use of this material is permitted.
  Permission from IEEE must be obtained for all other uses, in any current or future 
  media, including reprinting/republishing this material for advertising or promotional 
  purposes, creating new collective works, for resale or redistribution to servers or 
  lists, or reuse of any copyrighted component of this work in other works. 
  DOI: \href{http://dx.doi.org/10.1109/LCOMM.2008.080372}{10.1109/LCOMM.2008.080372}
}
\newcommand\copyrightnotice{%
\begin{tikzpicture}[remember picture,overlay]
\node[anchor=south,yshift=10pt] at (current page.south) {\fbox{\parbox{\dimexpr\textwidth-\fboxsep-\fboxrule\relax}{\copyrighttext}}};
\end{tikzpicture}%
}

\begin{document}

\title{Achieving Fair Network Equilibria with Delay-based Congestion Control
  Algorithms}

\author{Miguel~Rodríguez-Pérez, Sergio~Herrería-Alonso, Manuel
  Fernández-Veiga,~\IEEEmembership{Member,~IEEE,} \\
  Andrés Suárez-González and~Cándido~López-García%
  \thanks{The authors are with the Telematics Engineering Dept., Univ.~of
    Vigo, 36310 Vigo, Spain. Tel.:+34~986~813459; fax:+34~986~812116; email:
    miguel@det.uvigo.es (M. Rodríguez-Pérez). This work was supported by the
    ``Ministerio de Educación y Ciencia'' through the project
    TSI2006-12507-C03-02 of the ``Plan Nacional de I+D+I'' (partly financed
    with FEDER funds).}}

\maketitle
\copyrightnotice
\begin{abstract}
  Delay-based congestion control algorithms provide higher throughput and
  stability than traditional loss-based AIMD algorithms, but they are
  inherently unfair against older connections when the queuing and the
  propagation delay cannot be measured accurately and independently. This
  paper presents a novel measurement algorithm whereby fairness between old
  and new connections is preserved. The algorithm does not modify the dynamics
  of congestion control, and runs entirely in the server host using locally
  available information.
\end{abstract}

\begin{keywords}
  Delay-based congestion control, FAST TCP, persistent congestion, fairness.
\end{keywords}

\IEEEpeerreviewmaketitle

\section{Introduction}

\PARstart{D}{elay-based} congestion avoidance (DCA) algorithms, such as FAST
or Vegas, achieve high throughput in high-speed long-latency
networks~\cite{wei06,brakmo94}. But it is also well known that their
equilibrium transmission rates are very sensitive both to the accuracy of the
estimated round-trip propagation delay and to the estimated queuing delay.
Measurement errors in any of these quantities may lead to severe unfairness. A
situation like that arises, for instance, when a new flow encounters a state
where the queue ahead of the bottleneck link never gets empty, thus hampering
to correctly estimate the propagation delay along its network path. This
harmful, self-sustained condition, termed \emph{persistent congestion}, was
already found as early as in~\cite{mo99}.

In~\cite{low02}, a mathematical analysis is provided for a scenario where
persistent congestion is due to the successive arrival of a set of everlasting
flows to an empty router queue. It has been argued that such scenario is far
unlikely, however, the arrival of just a single flow to a saturated link is a
sufficient condition to trigger unfairness as long as some of the older flows
do not depart. Such configuration, where a small group of newborn flows find a
link in equilibrium (bandwidth equally distributed) shared by $n$ preexisting
long-lived flows, was precisely the setting analyzed in~\cite{cui06}, and
includes~\cite{low02}, in fact, as a particular case. As a possible solution
to the persistent congestion problem,~\cite{cui06} suggests throttling down
briefly each newly started flow to allow queues to empty, and thus obtain a
reliable estimate of the propagation delay. We have found that this approach
is not always effective, though.

We show that such a cautious source can fail to measure a correct propagation
delay under general circumstances, and present a novel solution able to
remove the undesired effect of persistent congestion in arbitrary conditions.
As in~\cite{cui06}, our proposal only requires the modification of the sender
end host, and attains a throughput as high as (and a buffer utilization as low
as) FAST does.

\section{Equilibrium Rate of Recent Arrivals}
\label{sec:perscong}

Despite their differences at the packet level, all congestion control
algorithms can be mathematically described, at the flow level, by the
dynamical equation
\begin{equation}
  \label{eq:dynamical}
  \dot{w}_i(t) = \kappa_i(t) \left( 1 - \frac{q_i(t)}{u_i(t)} \right)
\end{equation}
where $w_i(t)$ denotes the congestion window at time $t$ for flow $i$,
$\kappa_i(t)$ is a gain function, $u_i(t)$ is a suitable utility function, and
$q_i(t)$ is the congestion signal~\cite{wei06}. The transmission rate is then
given by $x_i(t) = w_i(t) / r_i(t)$, where $r_i(t)$ is the round-trip time.
For DCA algorithms, $q_i(t)$ is the queuing delay. TCP Vegas uses $\kappa_i(t)
= 1 / r_i(t)$, whereas FAST takes $\kappa_i(t) = \gamma \alpha / \tau$,
where $\gamma$, $\alpha$ and $\tau$ are protocol parameters. Both instances,
FAST and Vegas, use $u_i(t) = \alpha / x_i(t)$ and have therefore equal
equilibrium structure, determined by~\eqref{eq:dynamical}, namely
\begin{equation}
    x^*_i = \frac{\alpha}{r^*_i - \hat{d}_i},
    \label{eq:xequil}
\end{equation}
where $\hat{d}_i$ is the propagation delay as estimated by flow $i$. 

We consider in this paper the arrival of a \emph{single} new flow (indexed by
$0$) at a bottleneck link of capacity $C$ shared by a set $\mathcal{F} = \{1,
\ldots, n\}$ of FAST flows. We also assume that each connection $f \in
\mathcal{F}$ knows its true round-trip propagation delay ($\hat{d}_f =
d_f$).\footnote{That is, we assume that there is a working algorithm in place
  to account for the persistent congestion bias. In Section~\ref{sec:solution}
  we present such an algorithm.} Hence, each flow $f$ is receiving $C / n$
units of bandwidth. Following the model in~\cite{cui06}, flow $f$ contributes
$\alpha$ packets to the router queues, so flow $0$ sees a propagation delay of
$\hat{d}_0 = d_0 + n\alpha/C$. As a result of this overestimated value, it
grabs a rate in the equilibrium
\begin{equation}
  x^*_0 = \frac{\alpha}{r_0^* - \hat d_0} = \frac{\alpha C}{b^*_0},
  \label{eq:x0}
\end{equation}
while the new common equilibrium rate for the older flows is
\begin{equation}
  x^*_f = \frac{\alpha C}{b^*_0+n\alpha}.
   \label{eq:xf}
\end{equation}
Since $\sum_{i=0}^n x_i^* = C$ we obtain
\begin{equation}
  \label{eq:b0}
  b^*_0 = \frac{\alpha}{2} \left( 1+\sqrt{1+4n} \right).
\end{equation}

The transmission rates given by~\eqref{eq:x0} and~\eqref{eq:xf} are clearly
unfair in that the recent arrival obtains far more bandwidth than the
rest. Moreover, the unfairness worsens with the number of flows, $x^*_0 /
x^*_f \sim O(\sqrt{n})$.

We claim that the fair equilibrium is achievable using a slightly modified
procedure to measure the propagation delay (see Section~\ref{sec:solution}).
Hence, since the onset of persistent congestion can be completely avoided, any
new flow will find the bottleneck link capacity fully and equally shared among
the older ones, as long as their rates have stabilized during the time elapsed
from the last arrival. Consequently, there is no need to pose the case of
successive flow beginnings, as in~\cite{cui06}, and the assumption of a single
recent arrival does not entail loss of generality.

\section{The Rate Reduction Approach}
\label{sec:ratereduc}

The solution presented in~\cite{cui06} consists in restraining transiently the
transmission rate of a new flow by a given factor to allow router queues to
get eventually empty, thus giving new connections a chance to directly measure
the true round-trip propagation delay. Unfortunately, and despite of the
reduction on its rate, the new connection is not always able to detect queue
emptiness. Note that, as the new flow drains queues by reducing its own rate,
competing flows respond by increasing their rates. Hence, the new flow will
only obtain the true propagation delay if queues empty before existing flows
are aware of this event, that is, if the time required to empty the queues is
less than the RTT of the existing flows.

Let $B^* = b^*_0 + n\alpha$ be the total backlog buffered at the core of the
network in equilibrium. This backlog will be drained from the queue at a rate
equal to the bottleneck link capacity minus the sum of the transmission rates
of all active flows. In the most favorable case, the new connection will
completely pause its transmission ($x_0 = 0$). Then, if all the existing flows
$f \in \mathcal{F}$ experience the same propagation delay ($d_f = d$), and so
the same~RTT ($r^*_f = r^*$), the fairness condition becomes
\begin{equation}
  \frac{B^*}{C - \sum^{n}_{f=1} x^*_f} < r^* = d + \frac{b^*_0+n\alpha}{C}.
  \label{eq:faircond1}
\end{equation}
Finally, substituting~\eqref{eq:b0}, \eqref{eq:xf} and $B^*$
into~\eqref{eq:faircond1}, it follows that
\begin{equation}
  d > \frac{n \alpha \left( 1+ \sqrt {1+4n}\right)}{2C} = \frac{n b_0^*}{C}
  \label{eq:faircond2}
\end{equation}
Thus, the rate reduction method is only effective when the round-trip
propagation delay of competing flows exceeds the lower bound calculated
in~\eqref{eq:faircond2}. This lower bound scales as $O(n^{3/2})$ with the
number of active flows, preventing a sensible default for the duration of the
rate reduction.

\section{A Novel Solution}
\label{sec:solution}

We noticed that, when the newly arriving flow stabilizes, it can indirectly
obtain a good estimation of its actual round-trip propagation delay. As
already pointed in Section~\ref{sec:perscong}, the new flow overestimates its
propagation delay as $\hat{d}_0 = d_0 + n\alpha/C$. Since $\hat d_0$ and
$\alpha$ are known, it suffices to estimate $n/C$ to get the real $d_0$.

A good estimation of $n$ and $C$ can be obtained even if the router queues are
not completely empty. In fact, as we will show, it suffices to just indirectly
measure queue length variations after a short change of the transmission rate.
Let $r_{0}^*$ be the RTT of the tagged flow once it reaches a stable
throughput. If this connection modifies its transmission rate
$x_0^{\prime}=(1-\theta) x_0^*$, with $\theta < 1$, for a brief time
$t_{\epsilon}$ (of the same order as $r^*$, so that the rest of the flows do
not adjust their own transmission rates) it will measure a new RTT
$r_{0}^{\prime}$ when it resumes its transmission. Let $\Delta r_{0} = r_{0}^*
- r_{0}^{\prime}$. Under such circumstances
\begin{equation}
  \label{eq:depletion}
  C \Delta r_{0} = \left(C-\sum_{i=1}^n
    x_i^*-(1-\theta)x_0^*\right)t_{\epsilon} 
\end{equation}
Substituting~\eqref{eq:x0},~\eqref{eq:xf} and~\eqref{eq:b0}
in~\eqref{eq:depletion}, and solving for $n$ yields
\begin{equation}
  \label{eq:estimated_n}
  \hat n = \frac{\theta t_{\epsilon}}{\Delta r_0} \left(
    \frac{\theta t_{\epsilon}}{\Delta r_0} - 1
    \right).
\end{equation}
Now, using~\eqref{eq:estimated_n} and~\eqref{eq:x0} $\hat C =
\frac{\left(1+\sqrt{1+4 \hat n}\right)x_0^*}{2}$ and the correct propagation
delay can be adjusted as
\begin{equation*}
  \hat d_{0}^{\prime} = \hat d_{0}^* - \alpha \frac{\hat n}{\hat C}.
\end{equation*}

Note that using positives values for $\theta$ causes the queue to drain, and
it is possible to exhaust the backlog before the end of the measure. In that
case~\eqref{eq:depletion} no longer holds and the number of flows is
overestimated. To avoid it, it suffices to use small negative values for
$\theta$, causing the queueing delay to increase. Although for insufficiently
dimensioned buffers this may cause some packet drops, this condition can be
easily detected and avoided by using smaller values of $\theta$ in subsequent
measures.

\section{Performance Analysis}
\label{sec:simul}

\begin{figure}
  \centering{%
    \subfigure[Single bottleneck topology.]
    {\includegraphics[width=0.67\columnwidth]{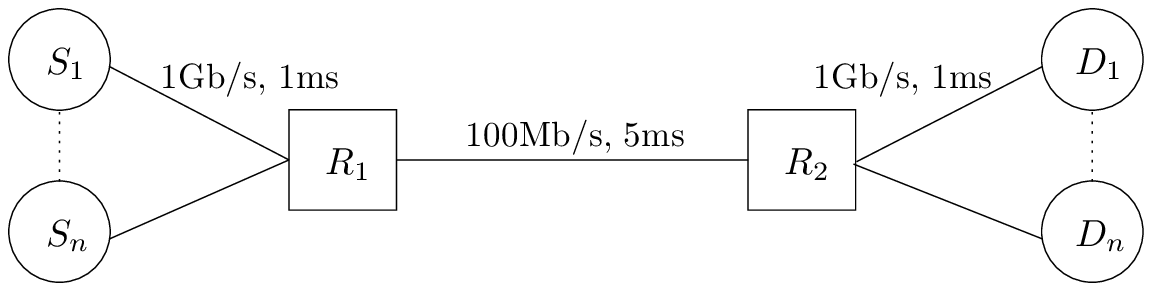}\label{net}}
    \subfigure[Multiple bottlenecks topology. Every link has a $100\,$Mb$/$s
    capacity and a $5\,$ms propagation delay.]
    {\includegraphics[width=0.8\columnwidth]{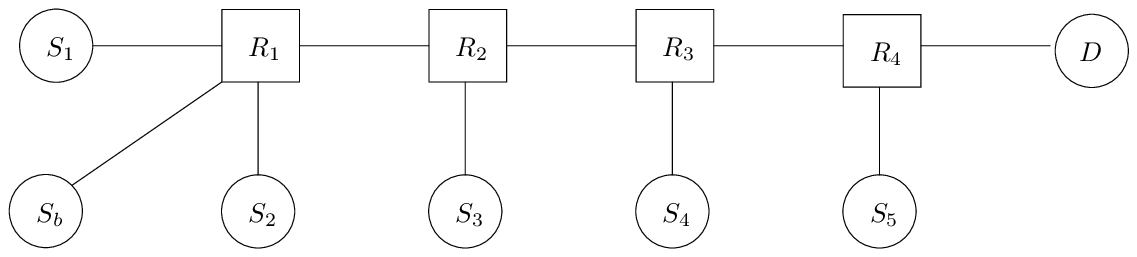}\label{net2}}}
  \caption{\label{graph} Network topologies used in the simulation
    experiments.}
\end{figure}

\begin{figure*}
  \centering
  \subfigure[Throughput comparison (FAST).]
  {\includegraphics[width=0.3\textwidth]{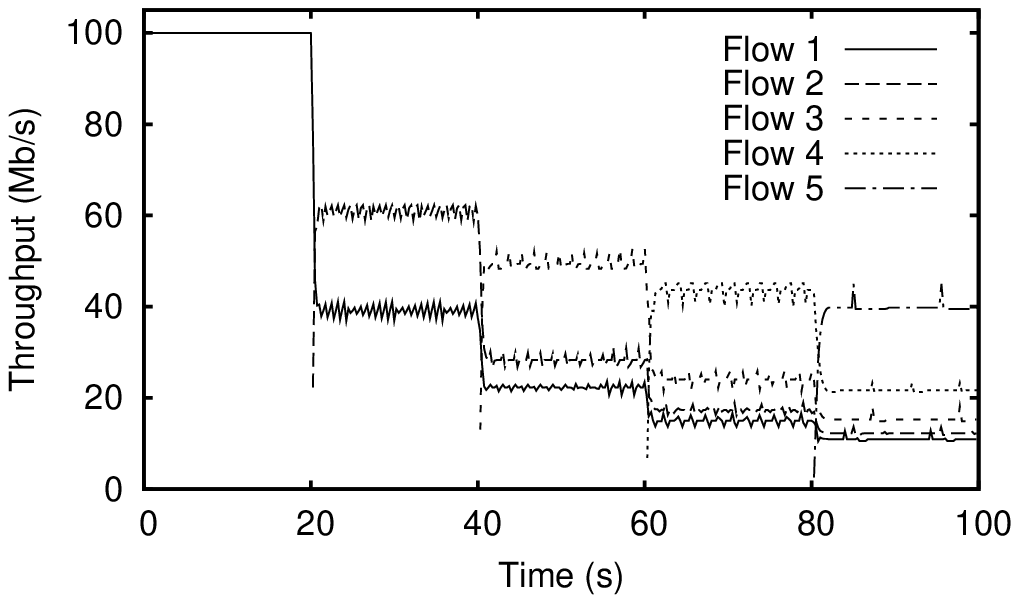} 
    \label{tputfast}}
  \subfigure[Throughput comparison (modified FAST).]
  {\includegraphics[width=0.3\textwidth]{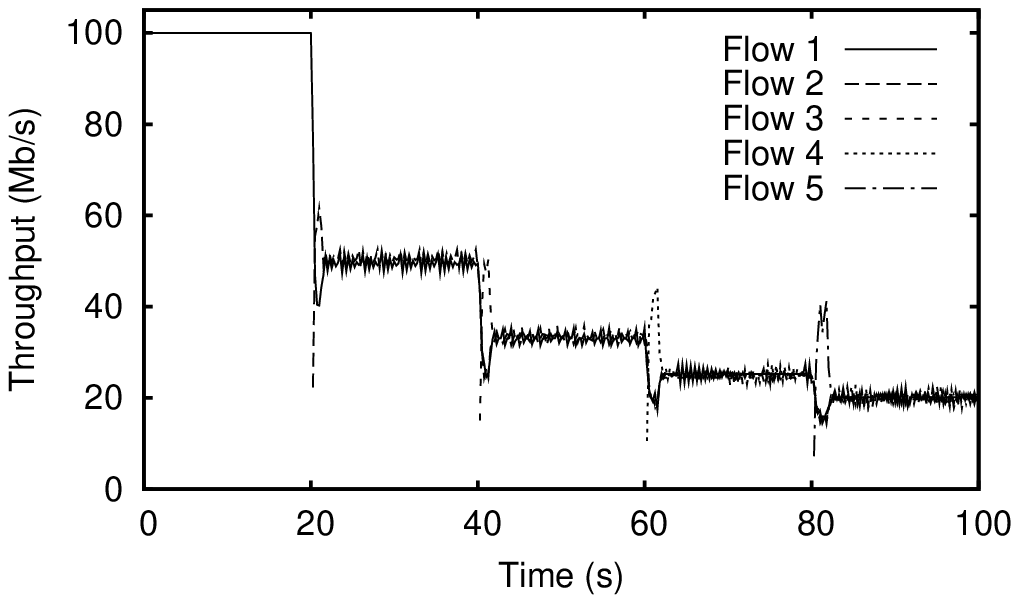} \label{tputenh}}
  \subfigure[Core queue length comparison.]
  {\includegraphics[width=0.3\textwidth]{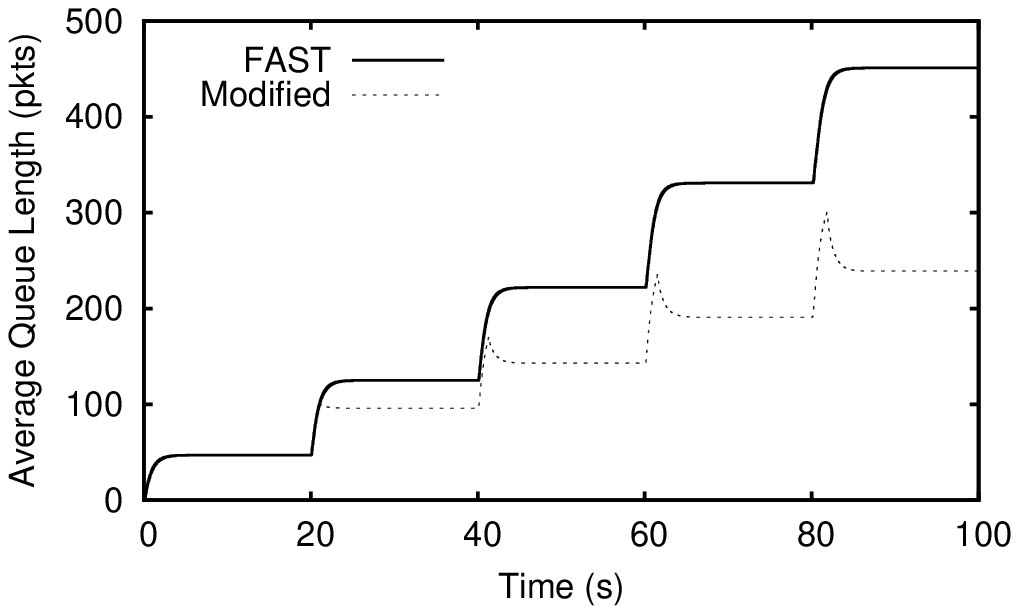} \label{cqueue}}
  \caption{Throughput and core queue length comparisons.}
\end{figure*}

\begin{figure*}
  \centering
  \subfigure[Impact of round-trip propagation delay.]
  {\includegraphics[width=0.3\textwidth]{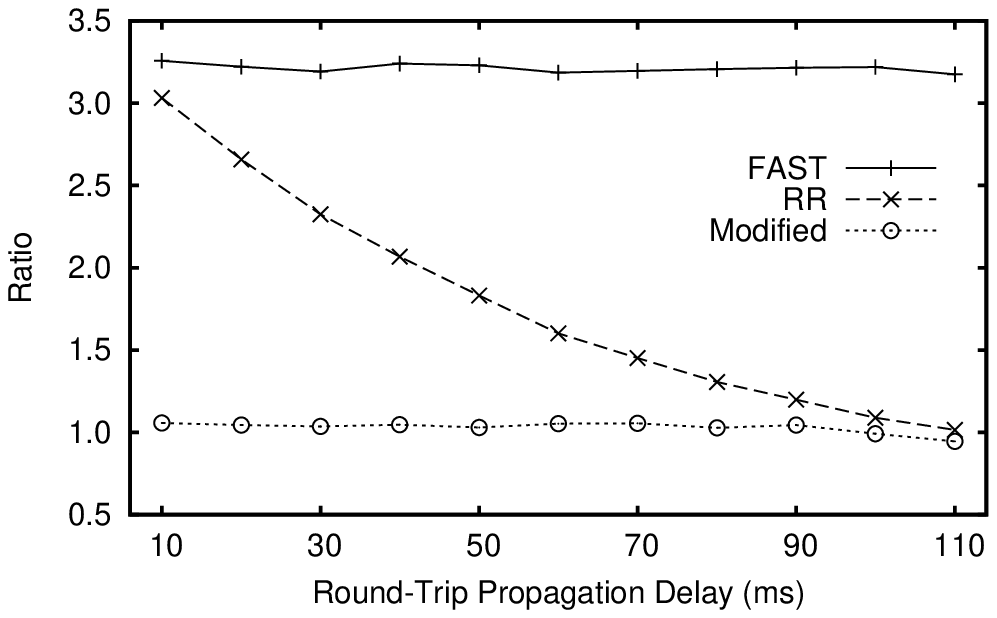} \label{rtt}}
  \subfigure[Impact of the number of preexisting flows.]
  {\includegraphics[width=0.3\textwidth]{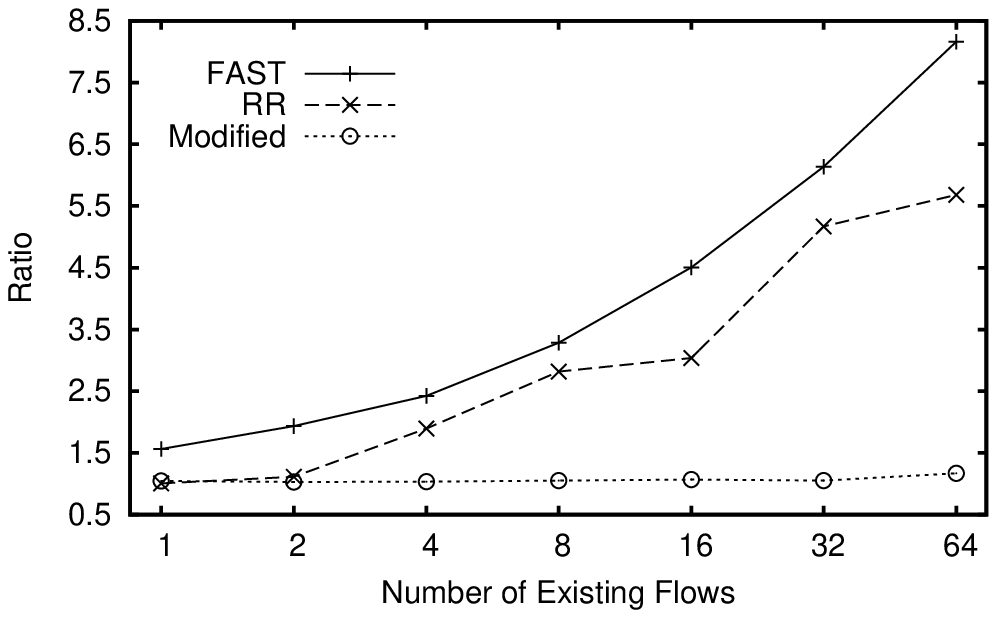} \label{nflows}}
  \subfigure[Impact of background traffic.]
  {\includegraphics[width=0.3\textwidth]{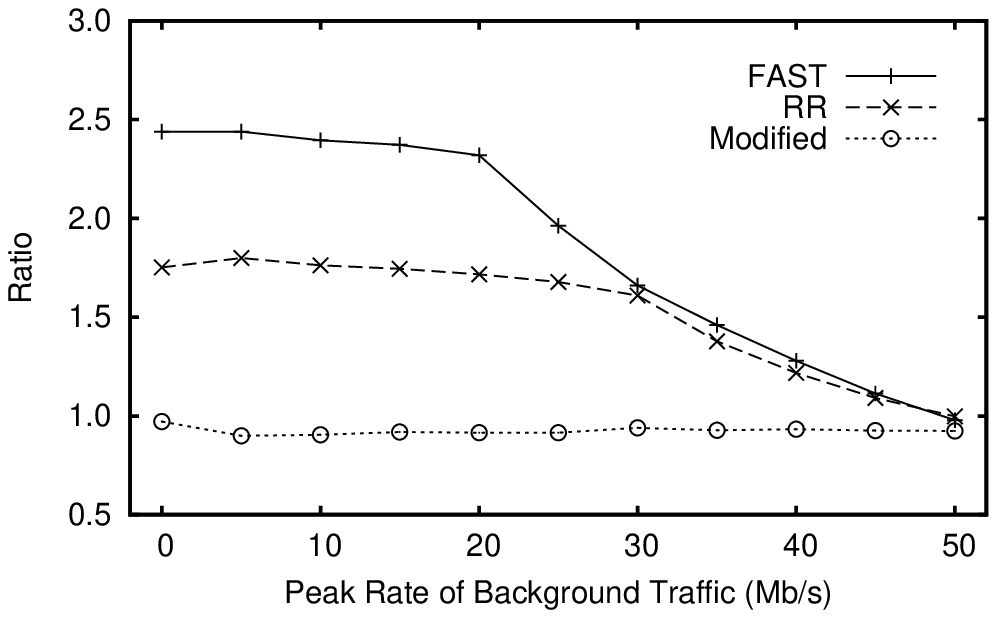} \label{back}}
  \caption{Simulation experiment results.}
\end{figure*}

To verify these claims, we report several ns-2 simulation experiments. In the
first one, there are five FAST connections $(S_i, D_i)$ sharing the bottleneck
link (Fig.~\ref{net}), starting at intervals of $20\,$s each. Routers' buffers
are large enough to avoid packet losses, and sources always have data to send.

Fig.~\ref{tputfast} shows the instantaneous throughputs of the FAST flows with
the original congestion avoidance mechanism ($\alpha = 50$ packets). As
expected, FAST strongly favors new sources and recent connections get larger
throughput than older flows. With the modified measurement method, this bias
disappears and the network bandwidth is shared fairly
(Fig.~\ref{tputenh}).\footnote{For the estimation of $\hat n$ we have employed
  $t_{\epsilon} = r^*$ and $\theta = -0.5$ to prevent the bottleneck from
  getting empty.} Also, the average queue length at the bottleneck
(Fig.~\ref{cqueue}) is consistently lower because, due to persistent
congestion, the backlog of FAST exceeds the target value of $\alpha$ packets
per source, whereas our proposal does not so.

A second test was run over the same network to compare the proposed algorithm
with the original FAST protocol and the rate reduction (RR) variant. Assume a
set of existing FAST flows aware of their true propagation delays, sharing the
bandwidth uniformly. Once their rates stabilize, a new flow starts. The delay
of the link $(R_1, R_2)$ was appropriately set so as to have the desired
RTT\@. Following customary practice, we measured the fairness among the new
and the $n$ existing connections as the ratio $n \bar{x}_0 / \sum^{n}_{f=1}
\bar{x}_f$ where $\bar{x}_0$ is the average transmission rate of the new flow
and $\bar{x}_f$ denotes the average rate of flow $f = 1, \ldots, n$.
Fig.~\ref{rtt} compares the performances of the three protocols for $n=8$. As
expected, with FAST, the new connection obtains a higher throughput. With the
RR method, the bandwidth sharing depends on the experienced propagation delay:
for delays below the threshold given by~\eqref{eq:faircond2} ($108\,$ms in
this scenario), the source rates become unfairer. In contrast, fairness is
preserved if the novel solution is used. Further, for any given RTT, the
unfairness aggravates with the number of flows, as Fig.~\ref{nflows} clearly
shows, either for FAST or for the RR reduction method, and only the modified
version allocates bandwidth equally.

A more realistic and stringent topology was also considered.
In~Fig.~\ref{net2}, the network (a variant of the classic parking-lot
topology) has multiple bottlenecks, with five flows running from nodes
$S_1,\ldots,S_5$ to node $D$. The flow originated in $S_1$ starts its
transmission after the rest of the flows stabilize. Additionally, in a similar
way as in~\cite{wei06}, some background traffic was simulated with a Pareto
flow $(S_b, D)$ with shape factor of $1.25$, average burst and idle time of
$100\,$ms and a peak rate ranging from $5$ to $50\,$Mb$/$s.  Fig.~\ref{back}
shows the results. Not surprisingly, with both FAST and the RR method,
fairness improves as the peak rate of background traffic increases.  The
reason is that, during active periods, FAST flows reduce their rates as router
queues fill due to background traffic, so in the idle periods the new flow can
seize a better estimate of its propagation delay before queues fill again. In
any case, our solution assures fairness irrespective of the amount of
background traffic introduced.

\section{Conclusions}
\label{sec:concl}

This paper has demonstrated that the \emph{rate reduction approach} fails to
solve persistent congestion in networks shared by many flows, as it cannot
always completely drain the bottleneck queues, and thus is unable to obtain an
accurate measure of the propagation delay.

We have presented a novel solution that does not rely on getting a direct
measure of the propagation delay. Instead, by carefully modulating its own
transmission rate, the source is able to calculate the error in the estimation
of the round trip propagation delay and thus share the link evenly with the
other FAST flows onwards.

\bibliographystyle{IEEEtran}
\bibliography{IEEEabrv,persistent}

\end{document}